\documentclass[12pt]{article}
\usepackage{epsfig}

\usepackage{amssymb}
\usepackage{amsmath}
\usepackage{amsfonts}
\usepackage{amsthm}

\theoremstyle{theorem}

\theoremstyle{definition}

\def\bp{\begin{proof}}
\def\ep{\end{proof}}

\def\be{\begin{equation}}
\def\ee{\end{equation}}
\def\ba{\begin{array}{c}}
\def\ea{\end{array}}

\def\ben{$$}
\def\een{$$}

\newcommand{\bea}{\begin{eqnarray}}
\newcommand{\eea}{\end{eqnarray}}

\newcommand{\bbr}{\br\!\br}
\newcommand{\kkt}{\kt\!\kt}

\newcommand{\kt}{\rangle}
\newcommand{\br}{\langle}
\begin{document}

\titlepage

\vspace{.35cm}

 \begin{center}{\Large \bf

Quantum inner-product metrics via recurrent solution of
Dieudonn\'{e} equation.

  }\end{center}

\vspace{10mm}

 \begin{center}

 {\bf Miloslav Znojil}

 \vspace{3mm}
Nuclear Physics Institute ASCR,

250 68 \v{R}e\v{z}, Czech Republic

{e-mail: znojil@ujf.cas.cz}

\vspace{3mm}

%\today, dreikronen.tex

\end{center}

\vspace{5mm}

%\newpage

\section*{Abstract}

A given Hamiltonian matrix $H$ with real spectrum is assumed
tridiagonal and non-Hermitian, $H\neq H^\dagger$. Its possible
Hermitizations via an amended, {\em ad hoc} inner-product metric
$\Theta=\Theta^\dagger>0$ are studied. Under certain reasonable
assumptions, {\em all } of these metrics $\Theta=\Theta(H)$ are
shown obtainable as recurrent solutions of the hidden Hermiticity
constraint $H^\dagger\ \Theta=\Theta\,H$ called Dieudonn\'{e}
equation. In this framework even the two-parametric
Jacobi-polynomial real and asymmetric $N-$site lattice
$H^{(N)}(\mu,\nu)$ is found exactly solvable at all $N$.

%\newpage

\section{Introduction \label{ch1}}

Given a set
 \be
 E_0^{(exp.)}\,,E_1^{(exp.)},\,\ldots\,,\,E_{N-1}^{(exp.)}
 \label{seta}
 \ee
of experimentally determined real eigenvalues of a quantum
observable $\mathfrak{h}$ (one may think, say, about excitation
energies of a heavy nucleus), we are often interested in its
simplified theoretical interpretation. Thus, typically \cite{Geyer},
one replaces the diagonalization of the realistic (and complicated)
Hamiltonian operator $\mathfrak{h}=\mathfrak{h}^\dagger$ by the
diagonalization of its simplified isospectral ``effective" image
 \be
 H=\Omega^{-1}\mathfrak{h}\,\Omega\, \neq H^\dagger\,.
 \label{dyson}
 \ee
In the similar half-phenomenological applications, some of the most
frequent sample sets (\ref{seta}) even carry convenient nicknames.
For example, if the levels are equidistant, one speaks about a
``vibrational spectrum". The reason is that we may always choose
then the alternative operator $H$ in a particularly simple form of
harmonic oscillator. For similar reasons, the name ``rotational
band" is being assigned to the non-equidistant sets of energies
$E_n^{(exp.)}$ which grow, with the level-numbering subscript $n$,
quadratically.

In the series of papers \cite{gegenb} - \cite{shendr} we proposed an
extension of the latter ``effective-operator" philosophy to a next
family of analytically tractable models where the values
(\ref{seta}) form a less elementary series but where they still
exhibit a certain regularity by being fitted, with a reasonable
precision, by a friendly $N-$plet
 \be
 E_0^{(theor.)}\,E_1^{(theor.)},\ldots,E_{N-1}^{(theor.)}
 \label{setth}
 \ee
defined as zeros of a suitable classical orthogonal polynomial of
degree $N$, $Y_N(E^{(theor.)}_j)=0$, $j =0, 1, \ldots, N-1$, cf.
Table \ref{pexp4}.

\begin{table}[ht]
\caption{Special tridiagonal Hamiltonians (\ref{kitiel}) related to
classical orthogonal polynomials
 \label{pexp4}}

\vspace{2mm}

\centering
\begin{tabular}{||c|c|c||c||c||}
\hline \hline
   \multicolumn{3}{||c||}{\rm matrix elements ($k=1,2,\ldots$)}
   &\multicolumn{1}{c||} {\rm polynomials}
   & {\rm ref.}      \\
    \hline
     $a_k=H_{k,k}$& $c_k=H_{k,k+1}$&$b_{k+1}=H_{k+1,k}$
    &\multicolumn{1}{c||}{$Y_k(E)\equiv |\psi\kt_k$}
     &
    \\
    \hline
 \hline
  $0$&$k/(2k+2a-2)$&$(k+2a-1)/(2k+2a)$&{\rm Gegenbauer}&\cite{gegenb}\\
  $2k+a-1$&$-k$&$-k-a$&{\rm Laguerre}&\cite{laguerre}\\
  $0$&$1+\delta_{k1}$&$1$&{\rm Tschebyshev}&\cite{sqw2}\\
  $0$&$1$&$2k$&{\rm Hermite}&\cite{hermit}\\
  $0$&$k/(2k-1)$&$k/(2k+1)$&{\rm Legendre}&\cite{shendr}\\
    \hline
 \hline
\end{tabular}
\end{table}

Formally, the exceptional character of the latter assumption
resulted from the reality, tridiagonality and asymmetry of the $N$
by $N$ matrix form of the underlying observable,
 \be
 H={H}^{(N)}=
  \left[ \begin {array}{cccccc}
   a_1&c_1&0&0&\ldots&0
  \\
  b_2
 &a_2&c_2&0&\ldots&0
 \\0&b_3&a_3&c_3&\ddots&\vdots
 \\0&0
 &\ddots&\ddots&\ddots&0
 %\\
 % {}\vdots&\ddots&\ddots
 %&b_{N-2}&a_{N-2}&c_{N-2}&0
 \\\vdots&\ddots&\ddots&b_{N-1}&a_{N-1}&c_{N-1}
 \\{}0&\ldots&0&0&b_{N}&a_{N}\\
 \end {array} \right]\,.
 \label{kitiel}
 \ee
In our present paper we intend to make the picture complete by
showing how the construction may be made feasible for {\em any}
classical orthogonal polynomial including the most complicated
two-parametric family of Jacobi polynomials for which our original
strategy of construction (as proposed in Ref.~\cite{gegenb} and used
in the other references of Table~\ref{pexp4}) failed.

Our present key idea will lie in the replacement of the
computer-assisted symbolic-manipulation method of Ref.~\cite{gegenb}
by its new alternative which proves virtually computer-independent.
In essence, we shall achieve our goal by making an explicit use of
the tridiagonality of our classical-polynomial-related matrices
$H^{(N)}$ and, moreover, of the positivity of their non-main
diagonals.

The presentation of our message will start, in section \ref{ch2}, by
the discussion of a few formal aspects of the situation in which the
observables of a given quantum system are represented by manifestly
non-Hermitian finite-dimensional matrices. In the subsequent section
\ref{ch3} we shall clarify the compatibility of such an approach
with the standard principles of Quantum Mechanics, recommending that
one should rather rewrite the currently used but slightly misleading
term ``non-Hermitian" as ``hiddenly Hermitian" {\it alias}
``cryptohermitian" \cite{SIGMA}. This hidden Hermiticity property of
$H$ is also characterized there by the so called Dieudonn\'e
equation which matches matrix $H$ with the so called metric, i.e.,
operator $\Theta=\Omega^\dagger\Omega \neq I$ related to
Eq.~(\ref{dyson}).

The key message of our present paper is delivered in section
\ref{ch4} where the Dieudonn\'e equation is shown solvable via
recurrences whenever the input matrix $H$ is assumed tridiagonal.
The case of Jacobi polynomials is then recalled for illustration -
this also completes the classical orthogonal polynomial list as
displayed in Table~\ref{pexp4}.

In the subsequent sections \ref{ch5} and \ref{ch6} we study, in more
detail, some consequences of a pentadiagonal-matrix choice of
$\Theta$ and $H$, respectively. One should add that while the former
results illustrate the merits of the present recurrent approach, the
latter section rather points at its natural limitations because, as
we shall show, not all pentadiagonal Hamiltonians $H \neq H^\dagger$
can be made cryptohermitian via a tridiagonal metric.

The last section \ref{ch7} is  summary.

\section{The doublets of Schr\"{o}dinger equations \label{ch2}}

Within the overall approach of Ref.~\cite{gegenb} we still have to
emphasize that the input quantities (\ref{seta}) and their
polynomial-zero fit (\ref{setth}) need not necessarily represent
just energies. In the full-real-line case of Ref.~\cite{hermit}, for
example, these values were treated as a discrete, non-equidistant
grid-point quantum representation of the observable $H \equiv Q$ of
a one-dimensional coordinate. This being said, we shall still speak
here, for the sake of brevity, just about a ``toy-model Hamiltonian
$H$".

Our present identification of the secular polynomial
$Y_N(E^{(theor.)})$ with one of the classical orthogonal polynomials
has a motivation in their simplicity. The recurrent method itself
remains applicable to a more general class of tridiagonal-matrix
models in Quantum Mechanics. In such a broadened perspective,
typically, one might like to fit a given sample of numerical values
of bound state energies via {\em any} prescribed full-matrix form of
$H$. In a preparatory step, we then merely have to employ the
Lanczos tridagonalization method \cite{Wilkinson} and convert the
given Hamiltonian into an infinite-dimensional tridiagonal matrix
${H}^{(\infty)}$. This matrix should further be truncated to yield
its $N-$dimensional version of the form (\ref{kitiel}). Indeed, once
our real and tridiagonal Hamiltonian matrix enters the linear
algebraic Schr\"{o}dinger equation
 \be
  H^{(N)}\,|\psi_n\kt=E_n\,|\psi_n\kt \,,\ \ \ \
 n = 0, 1, \ldots, N-1
  \,
 \label{SEnotime}
 \ee
any standard numerical technique might be used to solve it in
general case.

The specific advantage of our present preference of having
matrix~(\ref{kitiel}) generated by the recurrences between classical
orthogonal polynomials may be seen in the resulting immediate
knowledge of the eigenkets $|\psi_n\kt$ in closed form,
 \be
 |\psi_n\kt=\left (
  \ba
 \left (|\psi_n\kt\right )_1\\
 \left (|\psi_n\kt\right )_2\\
 \vdots\\
 \left (|\psi_n\kt\right )_{N-1}\\
 \left (|\psi_n\kt\right )_N
 \ea
 \right )=\left (
  \ba
 Y_0(E_n)\\
 Y_1(E_n)\\
 \vdots\\
 Y_{N-2}(E_n)\\
 Y_{N-1}(E_n)
 \ea
 \right )\,.
 \label{inputgen}
 \ee
At the same time, the main shortcoming of the approach may be
identified with the related manifest asymmetry (i.e., apparently,
non-Hermiticity) of our special matrices $H^{(N)}$  (for
illustration, cf. Table \ref{pexp4} once more). This asymmetry, in
particular, implies the necessity of an additional solution of the
conjugate linear algebraic Schr\"{o}dinger equation which determines
the {\em same} real eigenvalues but the entirely {\em different}
eigenvectors (which will be denoted, in what follows, as ``ketkets"
\cite{SIGMA}),
 \be
  \left [H^{(N)}\right ]^\dagger\,|\psi_m\kkt=E_m\,|\psi_m\kkt \,,\ \ \ \
 m = 0, 1, \ldots, N-1
  \,.
 \label{SEconj}
 \ee
The construction of ketkets is both important (i.a., it completes
the definition of a biorthogonal basis) and challenging. After all,
only in the simplest case of Tschebyshev polynomials of paper
\cite{sqw2} we were able to write down the explicit ketket analogue
of the explicit kets (\ref{inputgen}). We shall extend the latter
result in what follows.

\section{\label{ch3}Two alternative interpretations of ketkets}

In a way explained in \cite{SIGMA} and \cite{SIGMAdva} the most
important properties of  the ket solutions $|\psi_m\kt$ of
Eq.~(\ref{SEnotime}) and of the ketket solutions $|\psi_m\kkt$ of
Eq.~(\ref{SEconj}) should be seen in their mutual orthogonality {\it
alias} biorthogonality
 \be
 \bbr \psi_m|\psi_n\kt \neq  0 \ \ \ \ {\rm iff} \ \ \ m = n
 \ee
and in their biorthogonal-basis completeness,
 \be
 I =I^{(N)} = \sum_{n=0}^{N-1}\
 |\psi_n\kt\frac{1}{ \bbr \psi_n|\psi_n\kt}\
 \bbr \psi_n|\,.
 \label{upl}
 \ee
Their definition is ambiguous. As long as both Eqs.~(\ref{SEnotime})
and (\ref{SEconj}) are homogeneous, we may choose arbitrarily
normalized ``input" eigenvectors $|\psi_n^{[i]}\kt$ and
$|\psi_n^{[i]}\kkt$ with the real overlaps $\omega_n=\bbr
\psi_n^{[i]}|\psi_n^{[i]}\kt\neq 0$. Next, we may write down the
{\em general} solutions of Eqs.~(\ref{SEnotime}) and (\ref{SEconj})
in the respective rescaled forms
 \be
 |\psi_n\kt=|\psi_n^{[i]}\kt\times \alpha_n\,,\ \ \ \
 |\psi_n\kkt=|\psi_n^{[i]}\kkt\times \beta_n\,.
 \label{trafo}
 \ee
We will always choose the real proportionality constants $\alpha_n$
and $\beta_n$ in such a manner that $\bbr \psi_n|\psi_n\kt=1$, i.e.,
$I^{(N)} = \sum_{n=0}^{N-1}\
 |\psi_n\kt\,
 \bbr \psi_n|$. This means that we shall only
eliminate the rescaling ambiguity, say, in the kets,
 \be
 \alpha_n=\alpha_n(\beta_n)=
 \frac{1}{\beta_n\,\omega_n}\,.
 \label{reduce}
 \ee
The variability of $\beta_n$ (i.e., our freedom of rescaling the
ketkets) survives. From the point of view of its physical
interpretation, it may have two forms. They have to be discussed
separately.

\subsection{Fixed$-\mathfrak{h}$ scenario}

Whenever Eq.~(\ref{dyson}) contains a {\em given} non-unitary
operator $\Omega$ (so that also the metric becomes unique and
nontrivial, $\Theta=\Omega^\dagger\Omega \neq I$), the initial
Hermiticity property of $\mathfrak{h}=\mathfrak{h}^\dagger$ proves
equivalent to the easily derived cryptohermiticity relation
 \be
 \left [{H}^{(N)}\right ]^\dagger\,\Theta=\Theta\,{H}^{(N)}\,,
 \ \ \ \ \ \ \Theta:=\Omega^\dagger\Omega\,.
 \label{dieudo}
 \ee
We may re-write Eq.~(\ref{dieudo}) as an explicit definition of the
conjugate Hamiltonian $H^\dagger$. Once we assume the reality and
non-degeneracy of the spectrum of energies, the second
Schr\"{o}dinger Eq.~(\ref{SEconj}) may be compared with the first
one implying the proportionality rule
 \be
 |\psi_n\kkt=\gamma_n\,
 \Theta\,|\psi_n\kt\,.
 \label{ketkets}
 \ee
Constraint $\bbr \psi_n|\psi_n\kt=1$ eliminates the apparent new
freedom since
 \be
 \gamma_n=\gamma(\beta_n)=
 \frac{1}{\alpha_n^2(\beta_n)\,\br \psi_n^{[i]}|\Theta^{(N)}|\psi_n^{[i]}\kt}
 = \frac{\omega_n^2\,\beta_n^2}{\br \psi_n^{[i]}|\Theta^{(N)}|\psi_n^{[i]}\kt}
 \ee
where function $ \alpha_n=\alpha_n(\beta_n)$ was taken from
Eq.~(\ref{reduce}). We may conclude that even if the metric is given
in advance, the ketkets still remain ambiguous and may {\em vary}
with the $N-$plet of unconstrained rescaling parameters $\beta_n$.

\subsection{A variable$-\mathfrak{h}$ scenario \label{naserovy}}

Whenever we start from a given Hamiltonian $H^{(N)}$ and from a {\em
fixed} related biorthogonal basis, we may invert the above procedure
and reinterpret relations (\ref{dieudo}) as a set of constraints
imposed, by the requirement of the Hermiticity of a hypothetical
$\mathfrak{h}$, upon the metric $\Theta$. Relations (\ref{dieudo})
may be then called, for historical reasons, Dieudonn\'{e} equations
\cite{Dieudonne}. They will serve us as an implicit definition
determining the Hamiltonian-adapted and $\vec{\kappa}-$multiindexed
family of eligible metrics
$\Theta=\Theta^{(N)}=\Theta^{(N)}(H^{(N)},\vec{\kappa})$. Formally
we obtain the necessary explicit formula for them by multiplying
identity (\ref{upl}) by matrix $\Theta$ from the left. In the light
of Eq.~(\ref{ketkets}) this yields
 \be
 \Theta =\Theta^{(N)}(H^{(N)},\vec{\kappa}) = \sum_{n=0}^{N-1}\
 |\psi_n\kkt \ \kappa_n\
 \bbr \psi_n|\,,
 \ \ \ \ \ \ \kappa_n=
 \frac{1}{ \gamma_n}
 \,.
 \label{zohupl}
 \ee
This is a finite sum which defines the general $N-$parametric
solution of the Dieudonn\'{e} equation. One has to conclude that the
factorizations $\Theta =\Omega^\dagger\Omega$ will finally realize
the correspondence between our unique input Hamiltonian $H^{(N)}$
and its multiple eligible isospectral avatars
$\mathfrak{h}=\mathfrak{h}(\vec{\kappa})$.

In this context is is worth emphasizing that each of these avatars
may be accompanied by a few other operators $\mathfrak{g}$ of some
other observable quantities, with the corresponding (and now,
evidently, multiindex-dependent!) pullbacks $G^{(N)}(\vec{\kappa})$
accompanying the original, $\vec{\kappa}-$independent Hamiltonian
$H^{(N)}$. In this sense, the multiindex $\vec{\kappa}$ carries an
explicit additional physical information about the dynamical
contents of the model (cf. a more thorough discussion of this point
in \cite{Geyer}).

{\it Vice versa}, one may weaken or even remove these
multiindex-related ambiguities of the physical predictions via na
explicit assignment of an observable status to some of the members
of the family of operators $G^{(N)}(\vec{\kappa})$. The best known
example (and also, in parallel, one of the mathematically simplest
ones) is the requirement of the observability of a ``charge" as
conjectured by Bender, Boettcher and Jones \cite{BBJ}. In this
sense, the {\em band-matrix} (i.e., tridiagonal, pentadiagonal etc)
constructions of $\Theta$ (as used in papers listed in
Table~\ref{pexp4}) represent an {\em alternative} strategy which has
been proposed and supported by several phenomenological arguments in
Ref.~\cite{fund}.

\section{Recurrent constructions of the band-matrix metrics \label{ch4}}

\subsection{Dieudonn\'e equation and  diagonal metrics}

Relation~(\ref{dieudo}) will be treated here in the spirit of
paragraph \ref{naserovy}, i.e., as a linear set of algebraic
equations, with the input given by the matrix elements $a_1, b_1,
\ldots$ of our tridiagonal toy-model Hamiltonian (\ref{kitiel}), and
with the output giving the (non-unique \cite{Geyer}) definition of
the matrix elements of the eligible metrics $\Theta=\Theta(H)$.

As long as our Hamiltonian (\ref{kitiel}) is a real and tridiagonal
matrix, the simplest possible matrix form of the metric may be
assumed diagonal and real,
 \be
  \Theta= \left[ \begin {array}{ccccc}
  %\hline
   \theta_{1}&&&&
  \\
  %\hline
&\theta_{2}&&&
 \\
 &&\ddots&&
 \\
  &&&\theta_{N-1}&
 \\ &&&&\theta_{N}
 %\\
 %\hline
 \end {array} \right]\,.
 \label{kit0}
 \ee
By this ansatz the Dieudonn\'{e}'s equation gets converted into the
difference $\hat{Q}=\hat{H}^\dagger \Theta -
\hat{H}\,\Theta=\hat{Q}^\dagger=0$ between the real tridiagonal
matrix
 \be
 \left (\hat{H}^{(N)} \right )^\dagger\,\Theta=
  \left[ \begin {array}{cccccc}
  %\hline
   a_1^{}\theta_{1}&b_2^{}\theta_{2}&0&\ldots&0&0
  \\
  %\hline
  c_1^{}\theta_{1}
  &a_2^{}\theta_{2}&b_3^{}\theta_{3}&0&\ldots&0
 \\0&c_2^{}\theta_{2}&a_3^{}\theta_{3}&b_4^{}\theta_{4}&\ddots&\vdots
 \\\vdots&\ddots
 &\ddots&\ddots&\ddots&0
 \\{}0&\ldots&0&c_{N-2}^{}\theta_{N-2}
 &a_{N-1}^{}\theta_{N-1}&b_{N}^{}\theta_{N}
 \\{}0&\ldots&0&0&c_{N-1}^{}\theta_{N-1}&a_{N}^{}\theta_{N}\\
 %\hline
 \end {array} \right]\,
 \label{kitiellev}
 \ee
and its transposed partner
 \be
 \Theta\,\hat{H}^{(N)}=
  \left[ \begin {array}{cccccc}
  %\hline
   \theta_{1}a_1&\theta_{1}c_1&0&0&\ldots&0
  \\
  %\hline
 \theta_{2} b_2
&\theta_{2}a_2&\theta_{2}c_2&0&\ldots&0
 \\0&\theta_{3}b_3&\theta_{3}a_3&\theta_{3}c_3&\ddots&\vdots
 \\0&\ddots
 &\ddots&\ddots&\ddots&0
  \\{}0&\ldots&0&\theta_{N-1}b_{N-1}
  &\theta_{N-1}a_{N-1}&\theta_{N-1}c_{N-1}
 \\{}0&\ldots&0&0&\theta_{N}b_{N}&\theta_{N}a_{N}\\
 %\hline
 \end {array} \right]\,.
 \label{kitielle}
 \ee
The diagonal elements of this difference vanish so that we are left
with the single sequence of the recurrence relations
 \be
 \theta_{n+1}b_{n+1}=\theta_{n}c_{n}^{}\,,\ \ \ \ n = 1, 2,
 \ldots ,N-1\,.
 \label{recurv}
 \ee
We only have to verify that {\em all} of the matrix elements in
Eq.~\ (\ref{kit0}) remain strictly positive, $\theta_{j}>0$ for all
$j=1,2,\ldots,N$ \cite{Geyer}. In the diagonal-matrix case such a
test is trivial.

%\subsection{The assignment of a tridiagonal metric to the tridiagonal
%Hamiltonian}

\subsection{Dieudonn\'e equation and the  tridiagonal metrics}

Let us slightly simplify the conventional notation of
Eq.~(\ref{kitiel}) and redefine
 \be
 H={H}^{(N)}=
  \left[ \begin {array}{ccccc}
   a_{11}&a_{12}&0&0&\ldots
  \\
  a_{21}
 &a_{22}&a_{23}&0&\ldots
 \\0&a_{32}&a_{33}&a_{34}&\ddots
 \\0&0
 &a_{43}&a_{44}&\ddots
 \\\vdots&\ddots&\ddots&\ddots&\ddots
  %\\{}0&\ldots&0&0&a_{11}&a_{11}\\
 \end {array} \right]\,,
 \label{kikiri}
 \ee
postulating simply that $a_{jk}=0$ for $j>N$ or $k>N$. This will
simplify the formulae in the first nontrivial scenario in which we
replace the diagonal-matrix ansatz (\ref{kit0}) by its first
nontrivial, tridiagonal-metric alternative
 \be
 \Theta= \left[ \begin {array}{cccccc}
  {} b_{11}&{} b_{12}&0&0&\ldots&0
 \\{}{} b_{12}&{} b_{22}&{} b_{23}&0&\ddots&\vdots
 \\{}0&{} b_{23}&{} b_{33}&\ddots&\ddots&0
 \\0&0&\ddots&\ddots&{} b_{N-2{}N-1}&0
 \\ \vdots&\ddots&\ddots&{} b_{N-2{}N-1}&{} b_{N-1{}N-1}&{} b_{N-1{}N}
 \\0&\ldots&0&0&{} b_{N-1{}N}&{} b_{NN}\end {array}
 \right]\,.
 \label{tridii}
 \ee
The same recurrent approach as above may be applied when, {\it
mutatis mutandis}, we only replace Eqs.~(\ref{kitiellev}) and
(\ref{kitielle}) by their appropriate pentadiagonal-matrix
analogues.

The main diagonal of the difference $\hat{Q}$ vanishes so that the
discussion has to involve just the two separate diagonals of
$\hat{Q}$, viz., the outermost diagonal with
 \be
 \hat{Q}_{n{}n+2}=
 {} a_{{n+1}{n}}\,{} b_{{n+1}{n+2}}-{} b_{{n}{n+1}}\,{} a_{{n+1}{n+2}}
 =0
 \,,
  \ \ \ \ n=1,2,\ldots
  \label{forman}
  \,
 \ee
and the  remaining condition
 \be
 \hat{Q}_{n{}n+1}=
 {} a_{{n}{n}}\,{} b_{{n}{n+1}}+{} b_{{n+1}{n+1}}\,{} a_{{n+1}{n}}
 -{} b_{{n}{n}}\,{} a_{{n}{n+1}}
 -{} b_{{n}{n+1}}\,{} a_{{n+1}{n+1}}=0\,,
  \ \ \ \ n=1,2,\ldots
  \,
  \label{latman}
 \ee
In the first step let us turn attention to the former relations
(\ref{forman}) which define, in terms of an initial value, say,
$b_{{1}{2}}=1$, the sequence of the off-diagonal matrix elements in
the metric
 \ben
 b_{{n+1}{n+2}}=
 {\frac {{} b_{{n}{n+1}}\,{} a_{{n+1}{n+2}}}{{} a_{{n+1}{n}}}}\,
 \,,
  \ \ \ \ n=1,2,\ldots, N-2\,.
  \,.
  \,
 \een
In contrast to the brute-force symbolic-manipulation constructions
as sampled in full detail in Ref.~\cite{laguerre}, the occurrence of
an exceptional subscript $n_e=N$ such that $a_{{n_e+1}{n_e}}=0$ will
not make the construction more complicated. On the contrary, we get
an {\em easier} recurrence at $n=N-1$. Note also that an artificial
choice of an ``illegal" $b_{{1}{2}}= 0$ would merely return us back
to the previous formulae for diagonal metric.

In the second step let us remind the readers about the linearity of
the Dieudonn\'{e}'s equation. One might make use of the knowledge of
the diagonal-matrix solution $\Theta_{diagonal}$ and try to work
with an $\alpha-$dependent tridiagonal-metric ansatz
 \be
 \Theta=\Theta_{diagonal}+\alpha\,\Theta_{tridiagonal}\,.
 \label{tridiagm}
 \ee
In the nondiagonal component this would enable us to select
$b_{11}=0$  without any loss of generality. In this spirit our
second, remaining set of recurrences (\ref{latman}) degenerates to
the recurrent relations
 \ben
 b_{{n+1}{n+1}}=
 -{\frac {{} a_{{n}{n}}\,{} b_{{n}{n+1}}-{} b_{{n}{n}}\,{} a_{{n}{n+1}}
 -{} b_{{n}{n+1}}\,
 {} a_{{n+1}{n+1} }}{{} a_{{n+1}{n}}}}\,,
  \ \ \ \ n=1,2,\ldots,N-1
  \,
 \een
which define the diagonal matrix elements of $\Theta_{tridiagonal}$.
The algebraic part of our task is completed. One can easily verify
that our recurrences reproduce our older tridiagonal-metric results
derived by the symbolic-manipulation techniques and presented in the
references which are listed in Table~\ref{pexp4}.

A closer inspection of the Table reveals that  the
symbolic-manipulation-extrapolation techniques were only able to
cover the non-parametric cases (cf. the last three items 3. - 5.)
or, with certain much more serious but still tractable technical
difficulties, also the classical orthogonal polynomials with a
single variable parameter (cf. the first two items 1. - 2.).
Interested readers may find a more explicit description and/or a
deeper analysis of a suppression of these difficulties in
Ref.~\cite{gerdt}.

%\newpage
 \subsection{The ``missing example" of Jacobi polynomials}

The brute-force treatment of the ``missing" Jacobi's two-parametric
classical polynomials $Y_n(z) =P^{(\mu,\nu)}_N(z)$ seemed to be,
from the point of view of the extrapolation technique of
Ref.~\cite{gegenb}, prohibitively complicated. At the same time,
precisely these complications inspired our search for a new
approach.

From the point of view of applicability of our present recurrent
approach, there is virtually no difference between the simpler and
more complicated concrete forms of the input matrix of the
(tridiagonal) Hamiltonian. One only has to verify the assumptions.
In the case of Jacobi polynomials there emerge in fact no specific
problems. One just has to find the real and non-degenerate $N-$plet
$\{E_n\}$ of roots of the $N-$th Jacobi polynomial,
 \be
 P^{(\mu,\nu)}_N(E_n)=0\,,\ \ \ \ \ n = 0, 1, 2, \ldots, N-1\,.
  \label{seculare}
 \ee
For any practical purposes such a secular equation may be considered
``solvable" because the values of $E_n$ may be found not only by
means of standard numerical algorithms but also using their
combination with many available specific approximation formulae
\cite{Stegun}.

The presence of the two auxiliary free parameters $\mu,\nu
> -1$ may be expected to make the resulting spectrum sufficiently
flexible and, up to certain degree, sufficiently universal. Having
in mind a spectral-fitting applicability of formula
(\ref{seculare}), it makes sense to appreciate the compact
$N-$dimensional form of the kets
 \be
 |\psi_n\kt=\left (
  \ba
 \left (|\psi_n\kt\right )_1\\
 \left (|\psi_n\kt\right )_2\\
 \vdots\\
 \left (|\psi_n\kt\right )_{N-1}\\
 \left (|\psi_n\kt\right )_N
 \ea
 \right )=\left (
  \ba
 P^{(\mu,\nu)}_0(E_n)\\
 P^{(\mu,\nu)}_1(E_n)\\
 \vdots\\
 P^{(\mu,\nu)}_{N-2}(E_n)\\
 P^{(\mu,\nu)}_{N-1}(E_n)
 \ea
 \right )\,.
 \label{input}
 \ee
These ket-vectors may be further compactified using the {\it ad hoc}
change of parameters with $E_n-1:=\xi$, $\mu+k:=\mu_k$ and
$\mu+\nu+k:=\sigma_k$. This yields the ``optimal" parametrization
 \ben
 |\psi_n\kt=\left (
  \ba
 1\\
 \mu_1+\frac{1}{2}\,\sigma_2\,\xi
 \\
 \frac{1}{2}\,\mu_1\mu_2+\frac{1}{2}\,\mu_2\sigma_3\,\xi
 +\frac{1}{8}\,\sigma_3\sigma_4\,\xi^2
 \\
 \frac{1}{6}\,\mu_1\mu_2\mu_3+\frac{1}{4}\,\mu_2\mu_3\sigma_4\,\xi
 +\frac{1}{8}\,\mu_3\sigma_4\sigma_5\,\xi^2
 +\frac{1}{48}\,\sigma_4\sigma_5\sigma_6\,\xi^3
 \\
 \vdots\\
 \frac{1}{(N-1)!}\,\mu_1\mu_2\ldots \mu_{N-1}+
 \ldots
 +\frac{1}{(2N-2)!!}\,\sigma_{N}\sigma_{N+1}\ldots \sigma_{2N-2}\,\xi^{N-1}
 \ea
 \right )
 \,.
 \een
Obviously, it would not make too much sense to display here the
explicit forms of the matrix elements of the metrics. These formulae
would be just too long for the standard capacity of the printed
pages of a Journal. Still, due to their recurrent nature, the use of
standard software (e.g., of MAPLE) would still keep any needed
non-numerical or numerical manipulation with these elements feasible
and straightforward.

Having these constructions in mind (and introducing also one
additional abbreviation $\nu+k:=\nu_k$), we may make
Table~\ref{pexp4} complete by providing the explicit
Jacobi-polynomial-related definition of the three diagonals of
elements in matrix (\ref{kitiel}),
 \ben
 a_k=H_{k,k}=\frac{\mu^2-\nu^2}{\sigma_{2k-2}\sigma_{2k}}\,,\ \ \ \
 k = 1, 2, \ldots, N\,,
 \een
 \ben
 c_k=H_{k,k+1}=-2k\frac{\sigma_k}{\sigma_{2k-1}\sigma_{2k}}\,,\ \ \ \
 k = 1, 2, \ldots, N-1\,,
 \een
  \ben
 b_{k+1}=H_{k+1,k}=-2\frac{\mu_{k}\nu_k}{\sigma_{2k}\sigma_{2k+1}}\,,\ \ \ \
 k = 1, 2, \ldots, N-1\,.
 \een
This makes us able to convert the recurrences for Jacobi polynomials
into the twin Schr\"{o}dinger-type eigenvalue problem where our
candidate for the Hamiltonian is the real and tridiagonal matrix $H$
with the desirable properties of its outer diagonals.

We are quite close to the climax of the construction. In particular,
the resulting recurrences remain elementary in the diagonal-metric
case where we obtain
 \be
 \theta_{n+1}=\theta_{n}c_{n}^{}/b_{n+1}
 =\theta_{n}\,\frac{k\sigma_k\sigma_{2k+1}}{\mu_k\nu_k\sigma_{2k-1}}
 \,,\ \ \ \ n = 1, 2,
 \ldots ,N-1\,.
 \label{recurv}
 \ee
Similarly, one can proceed in the tridiagonal metric case where we
also just have to perform the insertions. For the above-mentioned
typographical reasons, these insertions will be left to the readers
as an exercise.

\section{Pentadiagonal metrics and the emergence of a
cut-off-dependence \label{ch5}}

With the pentadiagonal  real and symmetric-matrix ansatz for the
metric
 \be
 \Theta=\Theta_{diagonal}+\alpha\,\Theta_{tridiagonal}
 +\beta\,\Theta_{pentadiagonal}\,
 \label{pentadiagm}
 \ee
the only unknown matrix elements occur in its updated pentadiagonal
part
 \be
 \Theta= \left[ \begin {array}{cccccc}
  {} b_{11}&{} b_{12}&{} b_{13}&0&\ldots&0
 \\{}{} b_{12}&{} b_{22}&{} b_{23}&{} b_{24}&\ddots&\vdots
 \\{} b_{13}&{} b_{23}&{} b_{33}&\ddots&\ddots&0
 \\0&{} b_{2{}4}&\ddots&\ddots&{} b_{N-2{}N-1}&{} b_{N-2{}N}
 \\ \vdots&\ddots&\ddots&{} b_{N-2{}N-1}&{} b_{N-1{}N-1}&{} b_{N-1{}N}
 \\0&\ldots&0&{} b_{N-2{}N}&{} b_{N-1{}N}&{} b_{NN}\end {array}
 \right]\,
 \label{petidii}
 \ee
where, again, the convenient initializations $b_{13}=1$ and $b_{12}=
b_{11}=0$ may (though need not) be used.

The related heptadiagonal Dieudonn\'{e}'s equation
$\hat{Q}=\hat{H}^\dagger \Theta - \hat{H}\,\Theta=0$ is again just
an identity along its main diagonal, $\hat{Q}_{nn}=0$ at all
$n=1,2,\ldots$. The strictly three nontrivial series of conditions
have to be satisfied, therefore, viz., conditions
 \ben
 \hat{Q}_{n{}n+3}=0\,,\ \ \ \hat{Q}_{n{}n+2}=0\,,\ \
  \ \hat{Q}_{n{}n+1}=0\,,\ \ \ \ n = 1, 2, \ldots\,.
  \een
The first rule
 \ben
 \hat{Q}_{n{}n+3}= b_{{n+1}{n+3}}\, a_{{n+1}{n}}- b_{{n}{n+2}}\,
 a_{{n+2}{n+3}}=0\,,\ \ \ \ n = 1, 2, \ldots\,
 \een
degenerates again to the trivial recurrences for the outer diagonal
of the metric,
 \ben
 b_{{n+1}{n+3}}=
 {\frac { b_{{n}{n+2}}\, a_{{n+2}{n+3}}}{ a_{{n+1}{n}}}}
 \,,\ \ \ \ n = 1, 2, \ldots\,.
 \een
From the second rule
 \ben
 {} b_{{n}{n+2}}\,{} a_{{n}{n}}+{} a_{{n+1}{n}}\,{} b_{{n+1}{n+2}}-
 {} b_{{n}{n+1}}\,{} a_{{n+1}{n+2}}
 -{} b_{{n}{n+2}}\,{} a_{{n+2}{n+2}} =0
 \een
we extract the recurrences which specify the intermediate diagonal
of the metric,
 \ben
 {} b_{{n+1}{n+2}}=
 -{\frac {{} b_{{n}{n+2}}\,{} a_{{n}{n}}
 -{} b_{{n}{n+1}}\,{} a_{{n+1}{n+2}}-{} b_{{n}{n+2}}\,
 {} a_{{n+2}{n+2}
 }}{{} a_{{n+1}{n}}}}\,,\ \ \ \ n = 1, 2, \ldots\,.
 \een
Finally, the third rule
 \ben
 b_{{n-1}{n+1}}\,a_{{n-1}{n}}
 +a_{{n}{n}}\,b_{{n}{n+1}}+b_{{n+1}{n+1}}\,a_{{n+1}{n}}-
 \een
 \ben
 -
 b_{{n}{n}}\,a_{{n}{n+1}}-b_{{n}{n+1}}\,a_{{n+1}{n+1}}
 -b_{{n}{n+2}}\,a_{{n+2}{n+1}}=0\,,\ \ \ \ n = 1, 2, \ldots\,
 \een
(where we have to add the formal definition of $a_{01}=0$) implies
that we can readily evaluate also the remaining missing elements
lying along the main diagonal of the metric,
 \ben
 b_{{n+1}{n+1}}=
 -{\frac {b_{{n-1}{n+1}}\,a_{{n-1}{n}}+a_{{n}{n}}\,b_{{n}{n+1}}}{
 a_{{n+1}{n}}}}+
 \een
 \ben
 +
 {\frac {b_{{n}{n}}\,
 a_{{n}{n+1}}+
 b_{{n}{n+1}}\,a_{{n+1}{n+1}}+b_{{n}{n+2}}\,a_{{n+2}{n+1}}}{
 a_{{n+1}{n}}}}\,,\ \ \ \ n = 1, 2, \ldots\,.
 \een
An insertion of the concrete input matrix elements of the
Hamiltonian as listed in Table \ref{pexp4} confirms the
compatibility of our older results with these recurrences. It is
worth recalling, in particular, that the use of recurrent formulae
reveals why the older method encountered particular difficulties via
an emergence of the cutoff-dependence of some elements of the metric
in multidiagonal cases.

The pentadiagonal metrics offer the first illustration of this
phenomenon. One may recall, e.g., Ref.~\cite{laguerre} where an
irregularity has been detected in the cut-off dependent
pentadiagonal-metric ``ultimate" element $b_{{N}{N}}$. In the
present language, such an irregularity appears as a very natural
consequence of the truncation. Indeed, the related recurrent
definition at $n+1=N$ contains the component
$b_{{N-1}{N+1}}\,a_{{N+1}{N}}$ which vanishes ``anomalously", i.e.,
due to the truncation of the Hamiltonian $a_{{N+1}{N}}=0$ or,
equivalently, of the metric, $b_{{N-1}{N+1}}=0$.

Naturally, an extension of these considerations to the heptadiagonal
and higher band-matrix metrics is straightforward and may be left to
the readers.

Another remark must be added concerning the requirement of the
positivity of the metrics. As already noticed in the preceding
papers of this series, this is a more or less purely numerical
problem, and no news emerged when we simplified the algebraic
constructions of matrices $\Theta$.

\section{Pentadiagonal Hamiltonians and the tridiagonal
metrics \label{ch6}}

In place of a tridiagonal non-Hermitian $N$ by $N$ matrix
Hamiltonian $H^{(N)}$ of Eq.~(\ref{kitiel}) one might now feel
inclined to extend the scope of the method and to try to perform the
systematic recurrent reconstruction of the metrics
$\Theta=\Theta(H)$ (say, with one, three, five, \ldots nonvanishing
diagonals) in more general, multidiagonal-Hamiltonian perspective.

Unfortunately, the same approach leads to the new, less
user-friendly mathematical phenomena is such a case. Let us
therefore mention a few of them now. For the sake of definiteness
let us only concentrate on the case of the next-to-nearest-neighbor
interactions and assume that they are described by the following
pentadiagonal real Hamiltonian
 \ben
 H^{(N)}_{penta}=\left[ \begin {array}{ccccccc}
 {} a_{11}&{} a_{12}&{} a_{13}&0&0&\ldots&0\\
 {}{} a_{21}&{} a_{22}&{} a_{23}&{} a_{24}&0&\ldots&0\\
  {}{} a_{31}&{} a_{32}&{} a_{33}&{} a_{34}& \ddots&\ddots&\vdots\\
 {}0&{} a_{42}&\ddots&\ddots&\ddots &
 {} a_{N-3N-1}&0\\
 \vdots&\ddots&\ddots&{} a_{N-2N-3}&{} a_{N-2N-2}& {} a_{N-2N-1}&
 {} a_{N-2N}\\
 0&\ldots&0& a_{N-1N-3}&{} a_{N-1N-2}&{} a_{N-1N-1}&
 {} a_{N-1N}\\
 0&\ldots&0&0& {} a_{NN-2}&{} a_{NN-1}&{} a_{NN}\end {array}
 \right]\,.
 \een
For the sake of brevity let us only consider the tridiagonal ansatz
(\ref{tridii}) for the metric and list just a few most essential
consequences of the insertion of these ansatzs in Dieudonn\'{e}'s
Eq.~(\ref{dieudo}).

Firstly, in a complete parallel with our preceding considerations we
may rewrite Eq.~(\ref{dieudo}) as a condition imposed upon the
heptadiagonal matrix
$\hat{Q}=\hat{H}^\dagger\,\Theta-\Theta\,\hat{H}$ which is real and
symmetric.

The simplest condition concerns the outermost diagonals and reads
%
%vnejsi rekur.s
 \ben
 {} b_{n+2,n+3}\,{} a_{n+2,n}-{} b_{n,n+1}\,{} a_{n+1,n+3}=0\,,\ \ \ \
 n=1,2,\ldots\,.
 \een
This relation enables us to treat $ b_{12}$ and $ b_{23}$ as
arbitrary initial values of recurrences determining all of the
remaining off-diagonal matrix elements of the tridiagonal metric
$\Theta$,
 \ben
  b_{n+2,n+3}={\frac {b_{n,n+1}\,{} a_{n+1,n+3}}{a_{n+2,n}}}\,,
  \ \ \ \ n=1,3, 5,\ldots
  \ \ \ \ {\rm or}
  \ \ \ \ n=2,4, 6,\ldots
  \,.
 \een
In the similar manner the condition of annihilation of the next two
outermost diagonals of $\hat{Q}=\hat{Q}^\dagger$ reads
 \ben
 {} b_{{n+1}{n+2}}\,{} a_{{n+1}{n}}+{} b_{{n+2}{n+2}}\,
 {} a_{{n+2}{n}}-{} b_{{n}{n}}\,{} a_{{n}{n+2}}-
 {}  b_{{n}{n+1}}\,{} a_{{n+1}{n+2}} =0
 \een
and determines the sequence of the odd or even diagonal matrix
elements
 \ben
 b_{{n+2}{n+2}}={\frac {-{} b_{{n+1}{n+2}}\,{} a_{{n+1}{n}}
 +{} b_{{n}{n}}\,{} a_{{n}{n+2}}+{} b_{{n}{n+1}}\,{} a_{{n+1}{n+2}
 }}{{} a_{{n+2}{n}}}}\,, \ \ \ \ n = 1,2, \ldots
 \een
depending on the other pair of the arbitrary initial-value
quantities $ b_{11}$ and $ b_{22}$, respectively.

In the last step we are still left with the ``redundant" sequence of
the recurrences
%
%
%
%zbyva posledni rekur%
% \ben
% {} b_{34}\,{} a_{33}+{} b_{44}\,{} a_{43}+{} b_{45}\,{} a_{53}-
% {}  b_{23}\,{} a_{24}-{} b_{33}\,{} a_{34}-{} b_{34}\,{} a_{44} =0
% \een
%
%ktera urcuje kompatibilni diag. of Ham.
% \ben
% a_{44}=-{\frac {-{} b_{34}\,{} a_{33}-{} b_{44}\,{} a_{43}-{} b_{45}\,
% {} a_{53}+{} b_{23}\,{} a_{24}+{} b_{33}\,{} a_{34}}{{} b_{34}}}
% \een
 \ben
 {} b_{{n+1}{n+2}}\,{} a_{{n+1}{n+1}}+{} b_{{n+2}{n+2}}\,{} a_{{n+2}{n+1}}
 +{} b_{{n+2}{n+3}}\,{} a_{{n+3}{n+1}}-
 \een
 \ben
 -
 {}  b_{{n}{n+1}}\,{} a_{{n}{n+2}}-{} b_{{n+1}{n+1}}\,{} a_{{n+1}{n+2}}
 -{} b_{{n+1}{n+2}}\,{} a_{{n+2}{n+2}} =0\,,\ \ \ n = 0,1, 2,
 \ldots\,.
 \een
In contrast to the tridiagonal-Hamiltonian results we are now merely
allowed to set $b_{-1,0}=0$ at $n=0$. Obviously, we are left with
the set of the formally superfluous but still perfectly valid
constraints. Thus, these conditions must be read as a restriction
imposed directly upon the input Hamiltonian itself. In this role
these recurrences read
 \ben
 a_{{n+2}{n+2}}= a_{{n+1}{n+1}}+{\frac {
        {} b_{{n+2}{n+2}}\,{} a_{{n+2}{n+1}}
        +{} b_{{n+2}{n+3}}\, {} a_{{n+3}{n+1}}
        }
        {{} b_{{n+1}{n+2}}}}-
 \een
 \ben
 \ \ \ \ \ \
 -{\frac {
 {} b_{{n}{n+1}}\,{} a_{{n}{n+2}}
 +{} b_{{n+1}{n+1}}\,{} a_{{n+1}{n+2}}}
 {{} b_{{n+1}{n+2}}}}\,,\ \ \ n = 0,1, 2, \ldots
 \een
and act, with the fifth arbitrary initial value $a_{11}$, as a mere
step by step recurrent definition of the diagonal matrix elements of
the ``admissible" or ``tridiagonally hermitizable" pentadiagonal
input Hamiltonian $H^{(N)}_{penta}$.

\section{Conclusions \label{ch7}}

We can summarize that it is now possible to declare the
classical-polynomial oscillators exactly solvable. In our present
paper, our recent proposal \cite{gegenb} of the simulation of a
given empirical  set of energy levels (or, alternatively, of an
$N-$plet of discrete experimental eigenvalues of any other quantum
observable) by the  $N-$plet of zeros of a suitable classical
orthogonal polynomial $Y_N(E)$ found its final formulation based on
the vector-structured, recurrent treatment of the matrix Dieudonn\'e
equation. The menu of possible sufficiently elementary
identifications of an $N-$plet of components of wave functions with
the polynomial sequences $Y_n(E)$, $n=0,1,\ldots,N-1$ has been made
complete.

Originally, our present study has been motivated by the failure of
the assignment of the metrics to the ``last missing"
Jacobi-polynomial oscillators by the extrapolation method of
Ref.~\cite{gegenb}. A new approach has been developed, in which the
three-term recurrence relations which are satisfied by the classical
orthogonal polynomials were redesigned as finding their main use in
Dieudonn\'e equation itself.

Although the existence of the three-term recurrence relations
satisfied by the classical orthogonal polynomials $Y_n(x)$ is a very
well known fact, the use of these polynomials in the role of
components of a wave function $|\psi\kt$ always seemed to be
hindered by the manifestly non-Hermitian appearance of the
Schr\"{o}dinger-equation resembling truncated versions of these
recurrences.

The situation started to be changing when people realized that the
verification of Hermiticity is in fact strongly dependent on our
selection of the inner product of wave functions. A true
breakthrough (and a slow acceptance of the new paradigm) emerged
when the manifestly non-Hermitian appearance of the imaginary cubic
potential has been recognized as ``reparable" (cf., e.g., the
details of the history as outlined in \cite{Carl}).

In such a context the appeal of the
classical-orthogonal-polynomial-related tridiagonal real and
asymmetric Hamiltonians $H^{(N)}$ appeared to be twofold. Firstly,
their finite-dimensional ``effective-matrix" form fitted very well
the {\em mathematical} requirement of having an exactly solvable
quantum model, without too many formal subtleties and with a
nontrivial metric given in closed form. Secondly, the {\em physical}
information happened to be carried in these models, in a not too
usual manner, by the {\em doublet} of operators, viz, by a {\em
given} $H$ and {\em deduced} $\Theta$.

The double appeal of our $H^{(N)}$s in the context of model-building
might find its further support in potential applications in the
domain of quantum lattices and chain models. Two  minor obstacles
are encountered and removed. Firstly, in the context of mathematics
one gets rid of the apparent non-Hermiticity of the natural
tridiagonal candidate $H \neq H^\dagger$ for the Hamiltonian via an
{\em ad hoc}, non-unitary (a.k.a. Dyson's) isospectral map $\Omega$.
Secondly, in the context of physics, an additional merit is that the
set of the eligible metrics is restricted to the band-matrix ones,
characterized by a partial preservation of the concept of the
localization of the lattice sites \cite{laguerre}.

The key technical advantage of the model has been found in the
tridiagonality of its Hamiltonians. The attempted move beyond the
family of classical polynomials (i.e., the attempted transition to
the pentadiagonal and further $H$s) has been found to lead to a loss
of simplicity. For the tridiagonal Hamiltonians, on the contrary, we
have got rid of the apparent non-Hermiticity of the most natural
tridiagonal candidates $H \neq H^\dagger$ for the Hamiltonians in
full generality. This was achieved via the use of the concept of a
non-unitary (a.k.a. Dyson's) isospectral map $\Omega$ connecting a
certain unknown operator $\mathfrak{h}$ (which is assumed Hermitian
but, presumably, prohibitively complicated) with our friendly and
explicitly diagonalizable matrix $H^{(N)}$. Moreover, the set of the
eligible metrics $\Theta=\Omega^\dagger\Omega$ has been successfully
ordered as starting from a short-ranged sub-hierarchy characterized
by the $(2k+1)-$diagonal matrix structure of $\Theta$ with $k\ll
[(N+1)/2]$. Thus, at the small $k=0,1,\ldots$, a partial return to
the concept of a ``smeared" locality of the lattice has been
achieved \cite{laguerre,hermit}.

In our present paper we stressed the importance of making the
assignment of a menu of metrics to a preselected Hamiltonian $H$
less dependent on the simplicity of the elements of $H$. In all of
our older papers on the subject (cf. their list in
Table~\ref{pexp4}) such a simplicity has been required as a
necessary condition of the applicability of the
truncation-extrapolation method as proposed (and successfully
illustrated) in Ref.~\cite{gegenb}. With the new method we were now
able to complete the list and incorporate also the ``last missing"
two-parametric Jacobi polynomials into the overall scheme.

The reasons of the success of the construction may be found in a
certain recurrent structure which has been found hidden in the $N$
by $N$ matrix form of the Dieudonn\'e equation. Unfortunately, the
one-directional recurrent nature of the construction $H \to
\Theta(H)$ is only preserved in the tridiagonal-Hamiltonian cases.
The situation becomes less favorable for the pentadiagonal and
higher generalizations of $H$ since one then must fine-tune the
compatibility of the metric $\Theta$ with Hamiltonian $H$ in a
selfconsistent manner.

\subsection*{Acknowledgments}

Work supported by the GA\v{C}R grant Nr. P203/11/1433, by the
M\v{S}MT ``Doppler Institute" project Nr. LC06002 and by the Inst.
Res. Plan AV0Z10480505.

\end{document}